\documentclass[altaffilletter,aps,nofootinbib,twocolumn,prd,eqsecnum,preprintnumbers,superscriptaddress,10pt,floatfix]{revtex4-1}
\pdfoutput=1
\usepackage{graphicx}
\usepackage{amsmath}
\usepackage{subfigure}
\usepackage{amssymb}
\usepackage{amsfonts}
\usepackage{mathtools}
\usepackage{amssymb}
\usepackage{enumerate}
\usepackage{xcolor}
\usepackage{bm}
\usepackage{mathrsfs}
\usepackage{epstopdf}
\usepackage{url}
\usepackage{footnote}
\usepackage{textcomp}
\usepackage{dsfont}
\usepackage{ulem}
\usepackage{hyperref}
\usepackage{enumerate}   
\usepackage{appendix}
\usepackage{textcomp}
\usepackage{tipa}

\makeatletter
\newcommand*{\rom}[1]{\expandafter\@slowromancap\romannumeral #1@}
\makeatother

\begin{document}

\title{Twisted doughnuts: Thick disk torus around equatorial asymmetric black hole}

\author{Che-Yu Chen}
\email{b97202056@gmail.com}
\affiliation{RIKEN iTHEMS, Wako, Saitama 351-0198, Japan}

\author{Eva Hackmann}
\email{eva.hackmann@zarm.uni-bremen.de}
\affiliation{Center of Applied Space Technology and Microgravity (ZARM), University of Bremen, Am Fallturm 2, 28359 Germany}
\affiliation{Gauss-Olbers Space Technology Transfer Center, University of Bremen, 28359 Bremen, Germany}

\author{Audrey Trova}
\email{audrey.trova@ikmail.com}
\affiliation{Center of Applied Space Technology and Microgravity (ZARM), University of Bremen, Am Fallturm 2, 28359 Germany}
\affiliation{Leibniz University of Hannover, Institute for Theoretical Physics, Appelstraße 2, 30167 Hannover, Germany}


\begin{abstract}
The Kerr black hole spacetime is symmetric with respect to a well-defined equatorial plane. When such a symmetry is broken, for instance, by some putative effects beyond general relativity, the Keplerian circular orbits around the black hole are distorted vertically away from the equatorial plane by an amount depending on the orbital radius. As a result, the Keplerian thin disk acquires a curved surface. In this work, we extend such results to thick tori configurations by considering non-self-gravitating Polish doughnut models. We show that due to the equatorial asymmetry of the spacetime, the centers and the cusps of tori are distorted away from the original equatorial plane toward the same direction as that experienced by the stable Keplerian orbits, and the entire tori configurations are twisted toward that direction as well. The shape of the distorted tori is demonstrated explicitly using a constant specific angular momentum profile $\ell(r,y)=\ell_0$ of the disk fluid. However, the result also applies to non-constant profiles of $\ell(r,y)$ generically in the sense that any asymmetric profile of $\ell(r,y)$ that attempts to produce a symmetric tori configuration either turns out to be ill-defined near the equatorial plane or suffers from fine-tuning issues. 
\end{abstract}

\maketitle

\newpage

\section{Introduction}

In General Relativity (GR), the Kerr hypothesis states that isolated black holes in our universe are described by the spacetime of the Kerr family \cite{Kerr:1963ud}. In the past decade, the developments of black hole observations, e.g., the direct detection of gravitational waves from binary black hole mergers \cite{LIGOScientific:2016aoc}, as well as the observations of supermassive black hole images \cite{EventHorizonTelescope:2019dse,EventHorizonTelescope:2022wkp}, have ushered in a new era in which probing the spacetime near such mysterious objects becomes possible. In the next few decades, testing the Kerr hypothesis through black hole observations may provide a promising opportunity to extend the limitations of our understanding of black holes and strong-gravity regimes \cite{Bambi:2011mj,Barack:2018yly}. In particular, these observations may eventually reveal whether black holes in our universe are really described by GR, or whether some new physics beyond GR has to be taken into account to describe these compact objects.

The Kerr spacetime contains a few inherent symmetries. The two most trivial ones are stationarity and axisymmetry, meaning that the entire geometry remains unchanged, no matter at which time $t$ or from which azimuthal angle $\varphi$ the observer is looking at the black hole. In addition to these two symmetries, the Kerr black hole has another obvious symmetry, that is, the $\mathbb{Z}_2$ symmetry, which means that there is a well-defined equatorial plane and the spacetime is symmetric with respect to this plane. Motivated by the attempt to test the Kerr hypothesis, one may consider testing the $\mathbb{Z}_2$ symmetry of black holes through observations.{\footnote{The possibility of testing other Kerr symmetries, such as integrability and circularity, through black hole observations has been discussed in Refs.~\cite{Destounis:2021mqv,Chen:2023gwm,Ghosh:2024het}.}} More explicitly, any persistent observed $\mathbb{Z}_2$ breaking in a regime consistent with stationarity could indicate the existence of new physics beyond GR, given that asymmetric environmental effects are properly accounted for.{\footnote{The $\mathbb{Z}_2$ symmetry of rotating black holes could be broken in several scenarios, such as in the theories in which parity-violating interactions exist \cite{Cardoso:2018ptl,Cano:2019ore,Cano:2022wwo,Tahara:2023pyg}, or for compact objects in string-inspired models \cite{Bianchi:2020bxa,Bianchi:2020miz,Bah:2021jno,Fransen:2022jtw}. In fact, if one relaxes the asymptotic flatness assumption, $\mathbb{Z}_2$ asymmetric black hole solutions can also be found in GR \cite{Newman:1963yy}, which can exhibit interesting phenomenology \cite{Wu:2023eml,Meng:2025jej} (see also Ref.~\cite{Kleihaus:2003df}).}} It is thus very crucial to look for possible special features that can appear \textit{if and only if} the $\mathbb{Z}_2$ symmetry of black holes is broken. 

However, it turns out that the identification of observational features that appear if and only if the black holes have no $\mathbb{Z}_2$ symmetry is a totally non-trivial task. Take the shadow image as an example. The shadow signature produced by a $\mathbb{Z}_2$ asymmetric black hole is highly sensitive to the Liouville integrability of geodesic dynamics around the black hole, i.e., it depends on whether the spacetime has non-trivial Killing tensors with rank-$2$ order \cite{Carter:1968rr} or even higher. In Refs.~\cite{Staelens:2023jgr,Cunha:2024gke}, it was shown that if $\mathbb{Z}_2$ symmetry is broken, the shadow critical curve,{\footnote{This terminology is adopted from Ref.~\cite{Gralla:2019drh}. The critical curve on the image plane corresponds to infinitely lensed images and carries only information about the black hole geometry itself.}} which is defined by the impact parameters of spherical photon orbits around the black hole, may appear to be vertically asymmetric on the image plane when the observer is at an exact edge-on inclination. However, it turns out that such a vertically asymmetric image feature only appears when the Liouville integrability of the photon geodesic is also broken, such that higher-order images develop chaotic features. In Refs.~\cite{Cunha:2018uzc,Chen:2020aix,Lima:2021las}, it was shown that if the Liouville integrability is preserved due to the existence of a non-trivial rank-$2$ Killing tensor, the shadow critical curves are always vertically symmetric for arbitrary observational inclinations, even if the black holes have no $\mathbb{Z}_2$ symmetry. In such cases, the amount of $\mathbb{Z}_2$ symmetry violation may alter the apparent size of the shadow critical curves \cite{Chen:2022lct}, but the size change of shadows is certainly not a feature that is special only to $\mathbb{Z}_2$ asymmetry of black holes \cite{Lima:2021las}.

Another possible candidate for such kind of unique features is associated with accretion disks. When black holes are not $\mathbb{Z}_2$ symmetric, the Keplerian circular orbits of massive particles are shifted vertically away from the standard equatorial plane \cite{Datta:2020axm,Chen:2021ryb}. The amount of shift depends on the orbital radius. As a result, the Keplerian thin disk, when considered a collection of Keplerian orbits with different radii, thus acquires a curved surface \cite{Chen:2021ryb} as opposed to a flat disk plane in the usual $\mathbb{Z}_2$ symmetric scenarios. For a near-edge-on observer, the curved disk surface may generate special features in its morphology and leave imprints on the line profiles of the disk \cite{Chen:2024hpw}. One natural question is, what would happen if one considers a more astrophysically relevant disk configuration, for instance, accretion disks beyond the Keplerian description or even the thick disk tori? The main aim of this paper is to extend the conclusion in Ref.~\cite{Chen:2021ryb} by considering thick disk tori around $\mathbb{Z}_2$ asymmetric black holes, and see how the torus structure may be affected by the lack of $\mathbb{Z}_2$ in the black hole geometry. 

The thick accretion disk models, commonly known as Polish doughnut, were originally introduced in Refs.~\cite{1976ApJ...207..962F,1978A&A....63..221A} in the late 1970s as stationary and pressure-supported fluid configurations orbiting compact objects. These models describe geometrically thick disks in which pressure gradients play a fundamental role, in contrast to thin disk models where the vertical pressure support is negligible. The first self-consistent relativistic constructions of thick disks were developed in standard general-relativistic spacetimes, primarily the Schwarzschild and Kerr metrics. In Refs.~\cite{1976ApJ...207..962F} and \cite{1978A&A....63..221A}, it was shown that equilibrium configurations can be obtained by assuming a perfect fluid with constant specific angular momentum. In this framework, the disk structure is governed by an effective potential whose equipotential surfaces determine the disk morphology and naturally allow for the presence of a cusp through which accretion can occur without invoking viscosity. Thick accretion disks have also been constructed in various non-standard spacetimes, including distorted black holes, the q-metric, parametrized models, and the spacetimes arising from modified theories of gravity or alternative compact-object models \cite{Chakraborty:2014eha,Teodoro:2021ezj,Faraji:2020tmv,Cassing:2023bpt}. In such contexts, deviations from the Kerr geometry modify the location of critical orbits, i.e., the cusps and centers, and the topology of the equipotential surfaces, leading to potentially observable differences in the disk structure. The influence of electromagnetic fields on thick disk equilibria has been investigated as well, including configurations involving magnetized disks or charged fluids \cite{Komissarov:2006nz,Kovar:2016kqh}. In the context of spacetimes that are asymmetric to the equatorial plane, thick disks have been constructed in the Taub-NUT metric \cite{Jefremov:2016dpi} and in the C-metric \cite{Faraji:2021mft}. In both cases, the accretion disks turned out to be asymmetric with respect to the equatorial plane as well. It can therefore be conjectured that this feature is a general signature of the broken $\mathbb{Z}_2$ symmetry. However, to our knowledge, only constant angular momentum profiles have been considered so far in this context in the literature. In particular, applying non-constant profiles of angular momentum of the disk fluids, in this work we will further address the question of whether accretion disks naturally become asymmetric with respect to the equatorial plane if the spacetime has no $\mathbb{Z}_2$ symmetry, or whether the disk distortion induced by the geometrical effects of black holes can be compensated somehow by asymmetric profiles of the angular momentum of disk fluids. Our results strongly suggest that the broken $\mathbb{Z}_2$ of black hole geometries could be a sufficient condition for asymmetric disk configurations.

The paper is organized as follows. In sec.~\ref{sec:th}, we briefly introduce the specific $\mathbb{Z}_2$ asymmetric black hole model we will be mainly considering in this work, and its properties. In sec.~\ref{sec:conl}, we discuss the Polish doughnut models with constant specific angular momentum, showing that the thick disk structures are asymmetric with respect to the equatorial plane. Then, in sec.~\ref{sec:nonconl}, we consider non-constant specific angular momenta, focusing specifically on the proof of the no-go theorem of having symmetric disk structures so long as the black hole spacetime breaks $\mathbb{Z}_2$. Finally, we conclude in sec.~\ref{sec:con}. 

\section{The No$\mathbb{Z}$ black hole metric}\label{sec:th}

In this paper, we consider the phenomenological model of the $\mathbb{Z}_2$ asymmetric black hole spacetime proposed in Refs.~\cite{Chen:2020aix,Chen:2021ryb,Chen:2022lct}, hereafter the No$\mathbb{Z}$ black hole spacetime, whose metric $g_{\mu\nu}$ can be expressed as
\begin{widetext}
\begin{align}
g_{tt}&=-1+\frac{2Mr\left(r^2+a^2y^2\right)}{\left(r^2+a^2y^2\right)^2+\left(r^2-2Mr+a^2y^2\right)\tilde\epsilon(y)}\,,\label{gtt}\\
g_{\varphi\varphi}&=\frac{\left(1-y^2\right)\left(r^2+a^2y^2+\tilde\epsilon(y)\right)\left[r^4+a^4y^2+r^2\left(a^2+a^2y^2+\tilde\epsilon(y)\right)+a^2\tilde\epsilon(y)+2Mr\left(a^2-a^2y^2-\tilde\epsilon(y)\right)\right]}{\left(r^2+a^2y^2\right)^2+\left(r^2-2Mr+a^2y^2\right)\tilde\epsilon(y)}\,,\label{gpp}\\
g_{t\varphi}&=-\frac{2Mra\left(1-y^2\right)\left(r^2+a^2y^2+\tilde\epsilon(y)\right)}{\left(r^2+a^2y^2\right)^2+\left(r^2-2Mr+a^2y^2\right)\tilde\epsilon(y)}\,,\label{gtp}\\
g_{rr}&=\frac{r^2+a^2y^2+\tilde\epsilon(y)}{r^2-2Mr+a^2}\,,\qquad g_{yy}=\frac{r^2+a^2y^2+\tilde\epsilon(y)}{1-y^2}\label{g23}\,,
\end{align}
\end{widetext}
in the Boyer-Lindquist coordinate system $(t,r,y,\varphi)$ with $y\equiv\cos\theta$. In addition to the mass $M$ and the spin $a$ of the black hole, the metric functions contain a deviation function $\tilde\epsilon(y)$. In this work, the deviation function is assumed to be $\tilde\epsilon(y)\equiv\epsilon May$ where $\epsilon$ is a dimensionless parameter \cite{Chen:2021ryb,Chen:2024hpw}. In this case, the Ricci scalar $\mathcal{R}$ and the Kretschmann  scalar $\mathcal{K}$ of the spacetime at large $r$ is 
\begin{align}
\mathcal{R}&=\frac{6a\epsilon M^2 y}{r^5}+\mathcal{O}\left(r^{-6}\right)\,,\nonumber\\
\mathcal{K}&=\frac{48 M^2}{r^6} + \frac{72 a\epsilon M^2 y}{r^7}+\mathcal{O}\left(r^{-8}\right)\,.
\end{align}
Therefore, the spacetime is asymptotically flat and reduces to the Kerr one as $r\rightarrow\infty$. In addition, when $a$ and $\epsilon$ are not zero, the $\mathbb{Z}_2$ symmetry is broken, i.e., the spacetime is not symmetric with respect to $y=0$. The event horizon radius is determined by the equation $r^2-2M r+a^2=0$, which is the same as the Kerr case \cite{Chen:2020aix}. Also, the surface $r^2+a^2y^2+\tilde\epsilon(y)=0$ is a singular surface. Assuming that the spacetime is everywhere nonsingular outside the event horizon for all $|a|/M\le 1$, we can only have $|\epsilon|\le2$ \cite{Chen:2021ryb}.   

The No$\mathbb{Z}$ black hole spacetime \eqref{gtt}-\eqref{g23} has several interesting properties on top of being asymptotically flat and $\mathbb{Z}_2$ asymmetric. First of all, there is no conical singularity along the spin axis even if $\mathbb{Z}_2$ symmetry is broken \cite{Chen:2024hpw}. This is distinct from the Taub-NUT metric and C-metric, in which the thick disk configurations around them have already been studied. Also, the spacetime, by its construction \cite{Chen:2020aix}, contains a hidden symmetry that preserves the Liouville integrability of its geodesic dynamics. Therefore, this No$\mathbb{Z}$ black hole model turns out to be suitable as a phenomenological model to look for implications or observational features that are special to the violation of $\mathbb{Z}_2$ symmetry. In Ref.~\cite{Chen:2021ryb}, it was shown explicitly that the circular Keplerian orbits around the No$\mathbb{Z}$ black hole are not located on the standard equatorial plane $y=0$. Instead, each of them is located at a plane that is parallel to the equatorial plane. Such a vertical shift of orbits with respect to the equatorial plane becomes less significant for orbits with larger radii due to the asymptotic flatness of the spacetime. As a result, the Keplerian thin accretion disk, which can be regarded as a collection of Keplerian circular orbits ranging from the innermost stable circular orbit to some large radius, acquires a curved surface. Such a curved structure of the Keplerian disk may generate special patterns in the line profiles of the disk morphology \cite{Chen:2024hpw}.

In the following two sections, we will extend the previous work by considering accretion thick tori and will investigate how the violation of $\mathbb{Z}_2$ symmetry of the black hole geometry could alter the structure of the torus configurations.

\section{Polish doughnut with constant specific angular momentum}\label{sec:conl}

In this work, we will consider non-self-gravitating thick tori to describe the configuration of a thick disk around the No$\mathbb{Z}$ black hole. Suppose the thick accretion disk is described by a torus of fluid undergoing circular motion around the rotation axis of the black hole. The four-velocity of the fluid can be described by
\begin{equation}
u^\mu=u^t\left(1,0,0,\Omega\right)\,,
\end{equation}
where $\Omega\equiv u^\varphi/u^t$. The specific angular momentum $\ell$ is defined as $\ell\equiv -u_\varphi/u_t$. Imposing the normalization condition for the fluid $u^\mu u_\mu=-1$, one obtains
\begin{equation}
u_tu^\varphi=\frac{-\Omega}{1-\ell\Omega}\,.\label{utuphi}
\end{equation}

The four-acceleration of the fluid can be expressed as
\begin{align}
a_\nu&\equiv u^\mu u_{\nu;\mu}\nonumber\\
&=u^\mu u_{\nu,\mu}+\frac{1}{2}{g^{\alpha\beta}}_{,\nu}u_\alpha u_\beta\nonumber\\
&=-u^\alpha u_{\alpha,\nu}=-u^tu_{t,\nu}-u^\varphi u_{\varphi,\nu}\nonumber\\
&=\partial_\nu\ln{|u_t|}-u_tu^\varphi\left(\frac{u_\varphi}{u_t}\right)_{,\nu}\,,\nonumber\\
&=\partial_\nu\ln{|u_t|}-\frac{\Omega}{1-\ell\Omega}\ell_{,\nu}\,,\label{fouracc}
\end{align}
where we have used several times the normalization condition $u^\mu u_\mu=-1$. Also, from the second line to the third line, the stationary and axisymmetry assumptions have been used, i.e., $u^\mu u_{\nu,\mu}=0$. Finally, to obtain the last line, we have used Eq.~\eqref{utuphi}.

We then assume that the fluid is described by a perfect fluid whose energy-momentum tensor is
\begin{equation}
T_{\mu\nu}=\rho h u_\mu u_\nu+pg_{\mu\nu}\,,
\end{equation}
with $\rho$, $h$, and $p$ expressing the rest-mass density, specific enthalpy, and pressure, respectively. From the conservation of the energy-momentum tensor
\begin{equation}
T^\nu_{\mu;\nu}=\partial_{\mu}p+\rho h a_\mu=0\,,
\end{equation}
one can get the fluid Euler equations:
\begin{equation}
-\frac{1}{\rho h}\partial_\mu p=\partial_\mu\ln{|u_t|}-\frac{\Omega}{1-\ell\Omega}\ell_{,\mu}\,,\label{euler}
\end{equation}
where we have used Eq.~\eqref{fouracc}. 

Assuming a constant specific angular momentum $\ell_0$, the last term on the right-hand side of Eq.~\eqref{euler} is zero. We then integrate Eq.~\eqref{euler} to get
\begin{align}
&\mathcal{W}(r,y)\equiv-\int\frac{dp}{\rho h}\nonumber\\=&\,\ln|u_t|=\ln{\left(\frac{g_{t\varphi}^2-g_{tt}g_{\varphi\varphi}}{g_{\varphi\varphi}+2\ell_0g_{t\varphi}+\ell_0^2g_{tt}}\right)^{\frac{1}{2}}}\,.\label{potentialW}
\end{align}
The isosurfaces of the effective potential $\mathcal{W}$ on the $(r,y)$-plane can dictate the torus structure of the thick disk. Note that one can set the integration constant that, in principle, should appear in Eq.~\eqref{potentialW} such that the potential is zero at the outermost surface of the torus.

\begin{figure}[t]
\centering
\includegraphics[width=230pt]{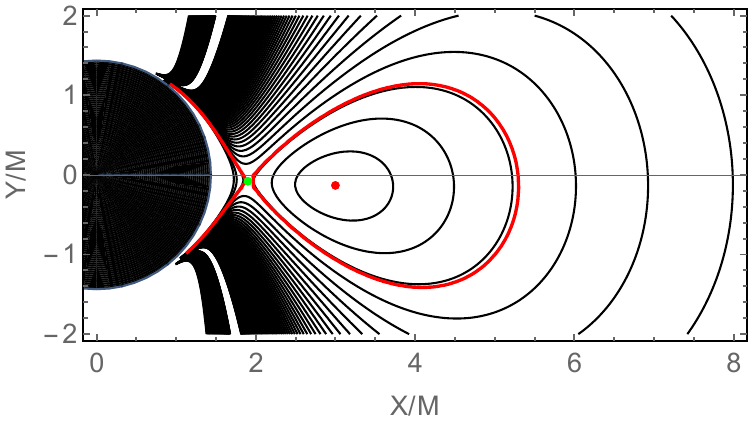}
\caption{The isosurfaces of the effective potential $\mathcal{W}$. Here we choose $a/M=0.9$, $\epsilon=2$, and $\ell_0=2.544M$. The red and green points represent the center and the cusp, respectively. The isosurface containing the cusp is highlighted by the red contour.}
\label{fig:potentialW}
\end{figure}

In Fig.~\ref{fig:potentialW}, we show the isosurfaces of the effective potential $\mathcal{W}$ in the Cartesian coordinates $(X\equiv r\sin\theta,\,Y\equiv r\cos\theta)$ for constant $\ell_0=2.544M$. In this figure, we choose $a/M=0.9$ and $\epsilon=2$. The two critical points, i.e., the center (red point) and the cusp (green point), are identified. They belong to the branches of stable and unstable Keplerian circular orbits, respectively. Both the center and the cusp are shifted downward with respect to the $y=0$ plane and are located at $(r/M,y)=(3,-0.046)$ and $(1.9,-0.038)$, respectively. 

\begin{figure}[t]
\centering
\includegraphics[width=230pt]{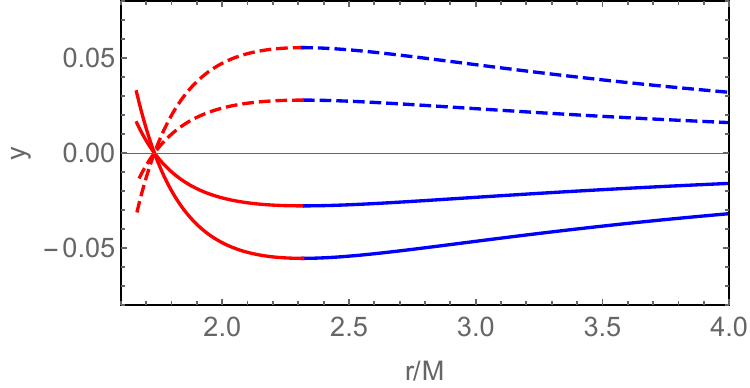}
\caption{The polar coordinate $y$ as a function of radius $r$ for stable (blue) and unstable (red) Keplerian circular orbits. The blue and red branches intersect at the innermost stable circular orbit (ISCO). We fix the spin value $a/M=0.9$ and consider various values of $\epsilon$. The curve whose ISCO is shifted further away from $y=0$ corresponds to $|\epsilon|=2$, and the other curve represents $|\epsilon|=1$. The solid (dashed) curves represent positive (negative) values of $\epsilon$.}
\label{fig:ryplot}
\end{figure}

Recalling that the No$\mathbb{Z}$ spacetime is Liouville integrable, the geodesic equations can be written in the first-order form. In particular, the radial and polar angle sectors of the geodesic equations are separable:
\begin{align}
&\left[r^2+a^2y^2+\epsilon May\right]^2(p^r)^2=\mathcal{R}(r)\,,\label{eqr}\\
&\left[r^2+a^2y^2+\epsilon May\right]^2(p^y)^2=\mathcal{Y}(y)\,,\label{eqy}
\end{align}
where $p^r$ and $p^y$ are the conjugate momenta associated with the $r$ and $y$ coordinates.{\footnote{More explicitly, we have $p^\mu=g^{\mu\nu}p_\nu$, where $p_\nu$ is the conjugate momenta of the coordinates $x^\nu$.}} The potentials on the right-hand side are defined by
\begin{align}
\mathcal{R}(r)\equiv&\,\left[E\left(r^2+a^2\right)-aL_z\right]^2\nonumber\\&-\left(\mathcal{K}+r^2\delta\right)\Delta-\left(L_z-aE\right)^2\Delta\,,\label{rpotential}\\
\mathcal{Y}(y)\equiv&\,\Big[\mathcal{K}+\left(L_z-aE\right)^2-a^2y^2\delta\nonumber\\-\epsilon May&\left(\delta-E^2\right)\Big]\left(1-y^2\right)-\left[a\left(1-y^2\right)E-L_z\right]^2\,,\label{ypotential}
\end{align}
where $E\equiv -p_t$, $L_z\equiv p_\varphi$, $\delta=1$ ($\delta=0$) for massive (massless) particles, $\Delta=r^2-2Mr+a^2$, and $\mathcal{K}$ is a separation constant. The Keplerian circular orbits are associated with the conditions $\mathcal{R}=d\mathcal{R}/dr=\mathcal{Y}=d\mathcal{Y}/dy=0$. In the absence of $\mathbb{Z}_2$ symmetry, the Keplerian circular orbits are shifted away from the $y=0$ surface as shown in Fig.~\ref{fig:ryplot}, where the stable and unstable branches of orbits are depicted in blue and red, respectively.

From Fig.~\ref{fig:ryplot}, one can see that in the No$\mathbb{Z}$ spacetime with a given set of $(a/M,\epsilon)$, the stable and some portions of unstable Keplerian orbits are shifted to opposite directions with respect to $y=0$. Note that the red curves actually cross the equatorial plane $y=0$ at a radius $r/M$ independent of the value of $\epsilon$ for a given $a/M$. This point corresponds to the marginally bound orbits in the Kerr spacetime, whose radius is denoted by $r_{mb}$. This interesting behavior can be understood by considering Eq.~\eqref{ypotential} with $\delta=1$. For marginally bound orbits with $E=1$, one can see that the entire potential $\mathcal{Y}(y)$ does not depend on $\epsilon$. Therefore, the marginally bound orbits are at the equatorial plane $y=0$ and degenerate in $\epsilon$. However, we have to emphasize that the property that the marginally bound orbits lie on the equator is highly related to the Liouville integrability of the No$\mathbb{Z}$ spacetime, and may not hold in other spacetimes whose $\mathbb{Z}_2$ symmetry is broken.

Although some portions of unstable Keplerian orbits are shifted to the opposite side of the equatorial plane with respect to stable orbits, the cusp in the torus is associated with the closed self-crossing equipotential surfaces, i.e., the red curve in Fig.~\ref{fig:potentialW}, and only appears on the same side of the equator as the center. In fact, the cusp radius is larger than $r_{mb}$ in this model, as we will demonstrate shortly. 

\begin{figure*}
    \centering
    \begin{tabular}{cc}
         \includegraphics[width=0.5\linewidth]{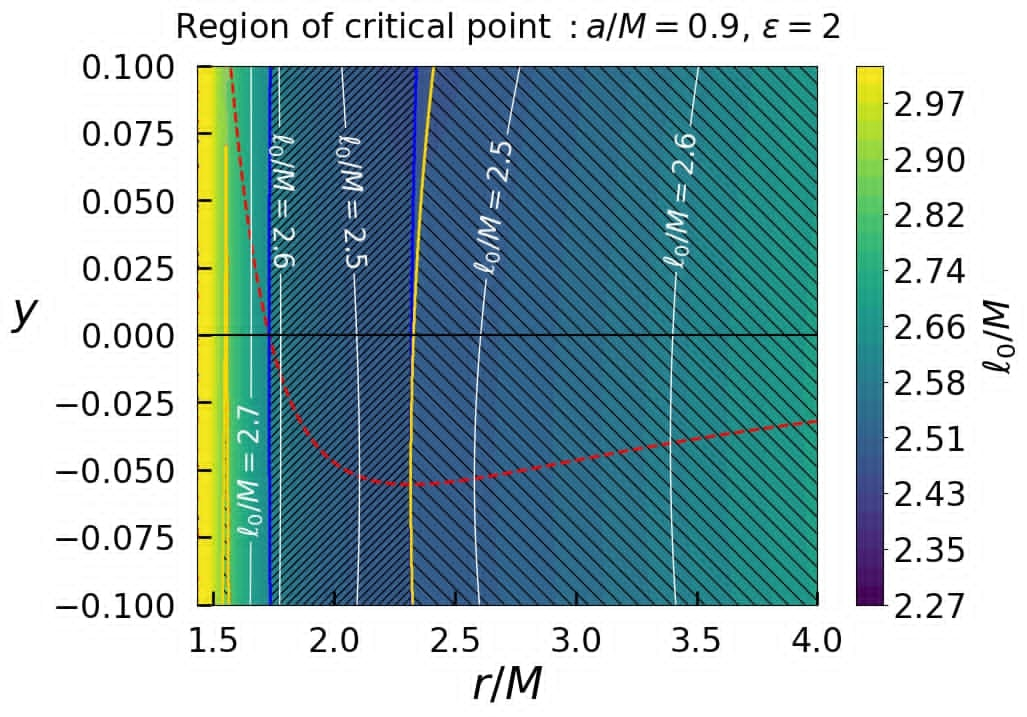}& \includegraphics[width=0.5\linewidth]{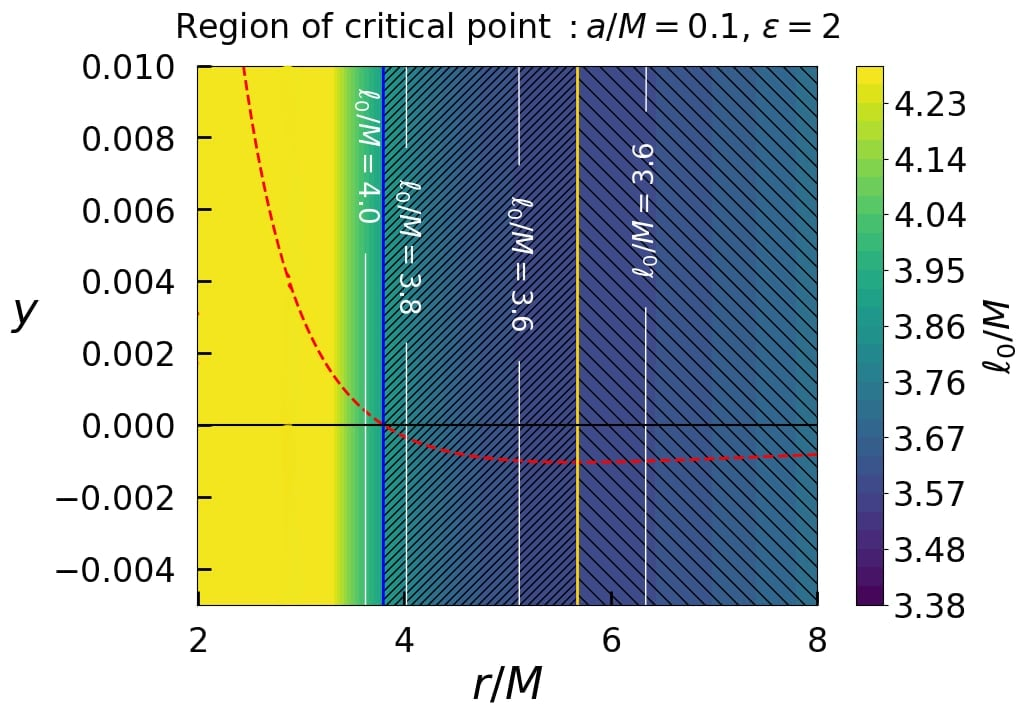} 
    \end{tabular}
    
    \caption{The region of critical points (cusp and center) in the $r$-$y$ plane is shown. These points must lie on the red dashed curve, which represents the locus of Keplerian circular orbits in the $r$-$y$ plane (see Fig.~\ref{fig:ryplot}). The critical points occur where $\partial_r \mathcal{W} = \partial_{\theta} \mathcal{W} = 0$, i.e., at the extrema of $\mathcal{W}$. The blue lines enclose a hatched region indicating where the mathematical conditions for a saddle point in $\mathcal{W}$ are satisfied, while the yellow lines mark the boundary of another hatched region where the conditions for a minimum in $\mathcal{W}$ are fulfilled. Finally, the color bar displays the values of $\ell(r,y)$ across the domain, and the white curve shows specific contours of constant $\ell_0$.}
\label{fig:critpoint}
\end{figure*}

The locations of the critical points (the cusp and the center) can be understood from Fig.~\ref{fig:critpoint}, which shows the region of the critical points that can appear around the black hole. In order to model an accretion disk, we focus on solutions that exhibit a family of bound, closed equipotential surfaces bounded by a self-crossing surface. The self-crossing surface represents the outer boundary of the disk and, as just mentioned, contains the accretion point (the cusp), while the center of the disk corresponds to the location where the pressure reaches its maximum, i.e., where the effective potential $\mathcal{W}$ attains a minimum. For such a configuration, the effective potential $\mathcal{W}$ must therefore possess a saddle point (the cusp) and a local minimum (the torus center). Fig.~\ref{fig:critpoint} summarizes the mathematical conditions required for these solutions, and we elaborate more as follows.

\begin{itemize}
    \item The extrema of $\mathcal{W}$, given by $\partial_r\mathcal{W}=0$ and $\partial_y\mathcal{W}=0$, lie on the red dashed line in Fig.~\ref{fig:critpoint}. Each point on the red dashed line represents a Keplerian circular orbit (see also Fig.~\ref{fig:ryplot}).
    \item The conditions required for a saddle point of $\mathcal{W}$ are indicated by the hatched region enclosed by the blue lines. In this region, the determinant of the Hessian matrix of $\mathcal{W}$ is negative.
    \item The region where $\mathcal{W}$ can exhibit a local minimum is shown by the hatched area enclosed by the yellow lines. In this region, the second derivative in the radial direction of ${\cal{W}}$ and the determinant of the Hessian matrix of $\mathcal{W}$ are positive.
\end{itemize}

As we have discussed, the marginally bound orbits are at the equatorial plane, as indicated by the point where the red dashed curves cross $y=0$. This point also coincides with one of the blue curves that represents the boundary of the allowed region for saddle points of $\mathcal{W}$. One sees that the segment of the red dashed curve contained in the hatched region bounded by the blue lines is on the same side as the center with respect to $y=0$. Therefore, cusp points and centers lie on the same side with respect to the equatorial plane. 

\begin{figure}[h]
\centering
\includegraphics[width=230pt]{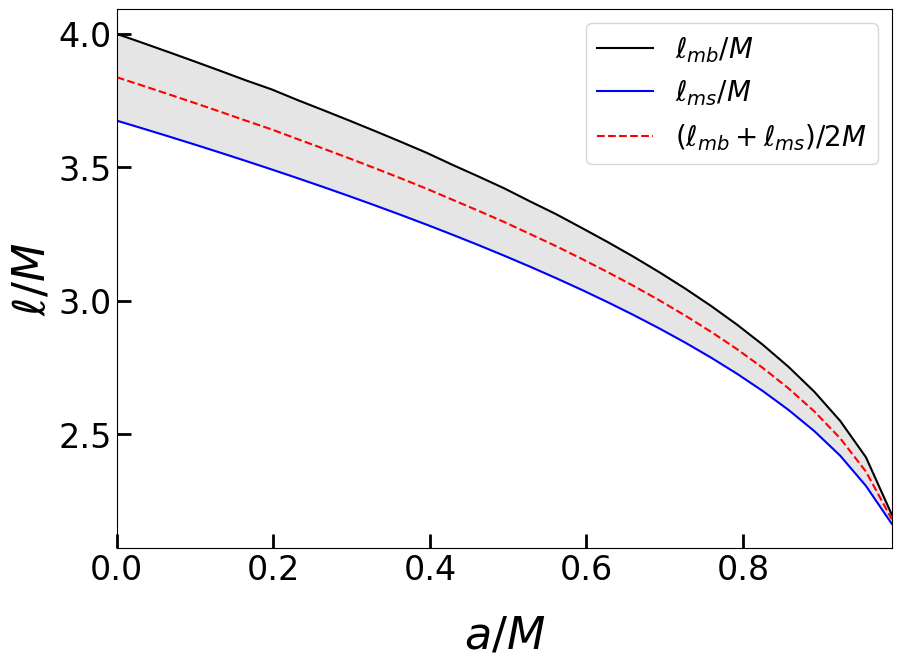}
\caption{The value of the specific angular momentum at the marginally bound orbit, $\ell_{mb}$, and the ISCO, $\ell_{ms}$, as functions of $a/M$ and $\epsilon = -2,0,2$. The dependence on $\epsilon$ is negligible over this range, and the corresponding curves are indistinguishable at the scale of the figure.}
\label{fig:lmblms}
\end{figure}

Furthermore, the color bar displays the range of $\ell_0$ values in these regions. These $\ell_0$ values correspond to the gray area shown in Fig.~\ref{fig:lmblms} for the chosen values of $a/M$ and $\epsilon$; that area denotes the range $\ell_{ms}<\ell_0<\ell_{mb}$ where bound solutions are possible. For instance, if one chooses $\ell_0=2.7M$ for $a/M=0.9$ and $\epsilon=2$, which is beyond the hatched region in the left panel of Fig.~\ref{fig:critpoint}, the effective potential $\mathcal{W}$ does not develop a cusp.

\begin{figure}
    \centering
    \includegraphics[width=230pt]{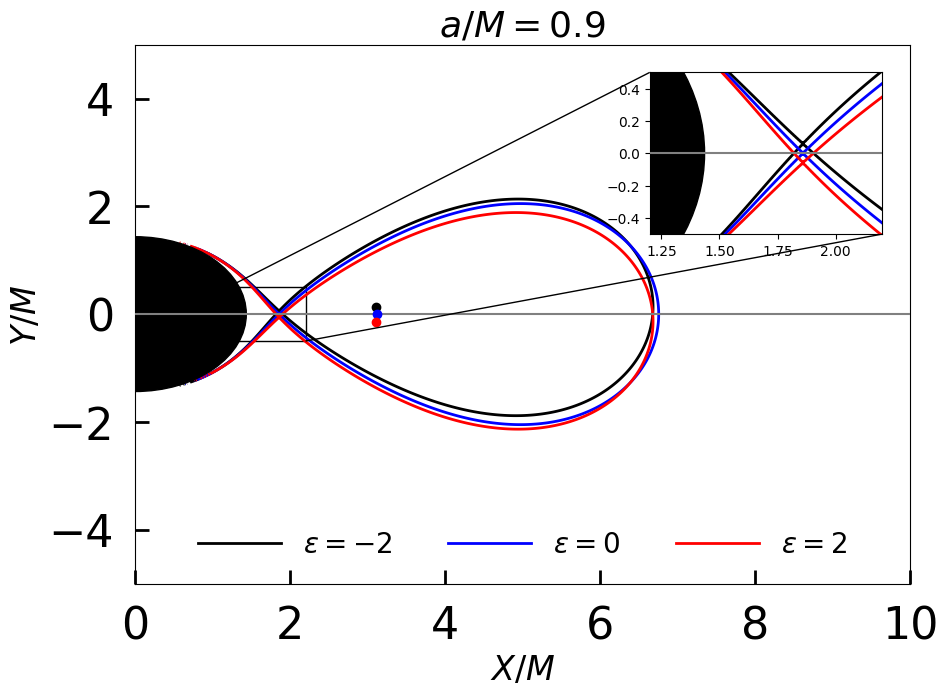}
    \caption{Equipotential contours of the effective potential $\cal{W}$ in the meridional plane for a fixed value of $a/M=0.9$. Each colored curve corresponds to the equipotential surface passing through the cusp of the disk for different values of $\epsilon$. The colored dots show the center of each disk. The inset highlights the region near the cusps in more detail.}
    \label{fig:MappEpsilon}
\end{figure}

\begin{figure}
    \centering
    \includegraphics[width=230pt]{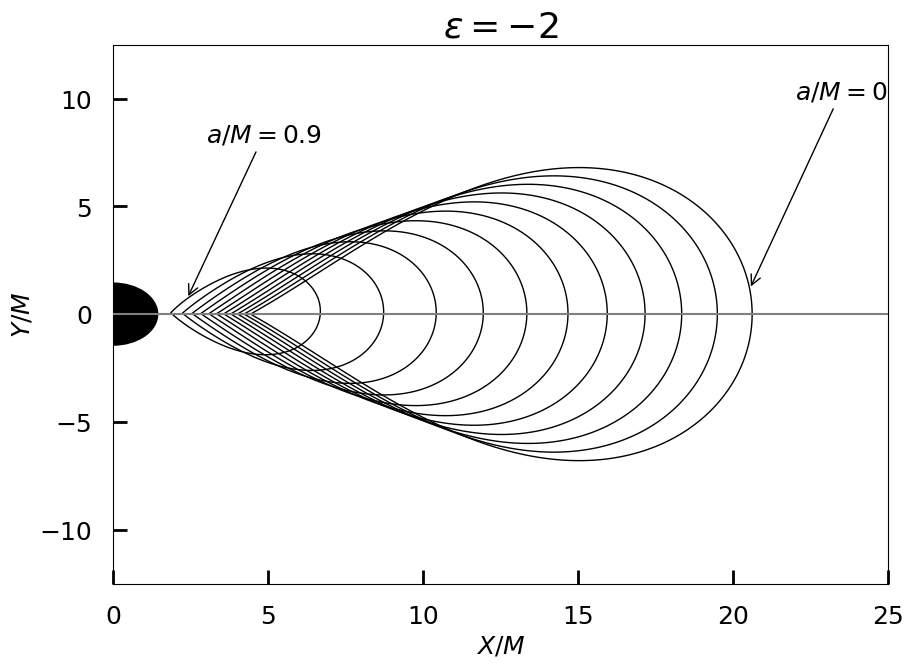}
    \caption{Equipotential contours of the effective potential $\cal{W}$ in the meridional plane for a fixed value of $\epsilon$. Each curve corresponds to the equipotential surface passing through the cusp of the disk for different values of $a/M$.}
    \label{fig:Mapa}
\end{figure}

To explore in more detail how the disk morphology changes with respect to the choice of the parameters $a$ and $\epsilon$ in the constant $\ell_0$ case, the values of $\ell_0$ have to be chosen with care. According to Fig.~\ref{fig:lmblms}, one can see that there is no unique value of the specific angular momentum $\ell_0$ that allows for bounded configurations with a cusp in the entire parameter space $(a/M,\epsilon)$ under our consideration. To systematically investigate the influence of the parameters $a$ and $\epsilon$ on the equipotential surfaces, we therefore fix $\ell_0 = \left(\ell_{mb} + \ell_{ms}\right)/2$ throughout the rest of the analysis in this section. The selected values of $\ell_0$ are indicated by the red dashed line in Fig.~\ref{fig:lmblms}, from which one can see its strong dependence on $a/M$ while the weak sensitivity to $\epsilon$. Fig~\ref{fig:MappEpsilon} displays the equipotential surface passing through the cusp for different values of $\epsilon$, while keeping the spin parameter fixed at $a/M = 0.9$. As expected, for a given value of $\epsilon$, both the center and the cusp lie on the same side of the equatorial plane: they are located below (above) the midplane for positive (negative) values of $\epsilon$ (see the inset of Fig.~\ref{fig:MappEpsilon}). This vertical asymmetry becomes more pronounced as the spin parameter $a/M$ increases. The effect of the spin parameter $a$ is illustrated in Fig.~\ref{fig:Mapa} for a fixed value of $\epsilon$. For clarity, we show only the equipotential surface originating from the cusp. As in the standard Kerr spacetime, increasing $a$ shifts both the center and the cusp inward. In addition, for larger values of $a$, the disk morphology becomes more bubble-like, while the inner region of the disk is noticeably thinner for small values of $a/M$.

\section{Non-constant specific angular momentum}\label{sec:nonconl}

In the case where the specific angular momentum of the disk fluid is not constant, i.e., $\ell=\ell(r,y)$, we consider the following equation
\begin{equation}
\partial_i\mathcal{W}=\frac{1}{2}g^{\alpha\beta}_{,i}u_\alpha u_\beta\,,\label{generalW}
\end{equation}
where $i=(r,y)$. Eq.~\eqref{generalW} is obtained by using the second line of Eq.~\eqref{fouracc} and considering only the circular orbits where $u_r=u_y=0$ such that the first term on the right-hand side vanishes. This equation can be further written explicitly as
\begin{equation}
\partial_i\mathcal{W}=\frac{u_t^2}{2}\left(\partial_ig^{tt}-2\ell\partial_ig^{t\varphi}+\ell^2\partial_ig^{\varphi\varphi}\right)\,,\label{generalW2}
\end{equation}
where the definition $\ell=-u_\varphi/u_t$ has been used.

Assume that the isosurfaces of the effective potential $\mathcal{W}$ are determined by curves $y(r)$. The contour of the curves satisfies $dy/dr=-\partial_r\mathcal{W}/\partial_y\mathcal{W}$. In general, for a given profile of $\ell(r,y)$, one can integrate the following equation
\begin{equation}
\frac{dy}{dr}=-\frac{\partial_rg^{tt}-2\ell\partial_rg^{t\varphi}+\ell^2\partial_rg^{\varphi\varphi}}{\partial_yg^{tt}-2\ell\partial_yg^{t\varphi}+\ell^2\partial_yg^{\varphi\varphi}}\,,\label{dydr}
\end{equation}
to obtain the isosurfaces $y(r)$. In our model with $\epsilon\ne0$ and for an arbitrary $\ell(r,y)$, the isosurfaces of $\mathcal{W}$ are asymmetric with respect to the standard equatorial plane $y=0$. It is natural to ask how general this statement is. In the following discussion, we will address the question of whether it is possible to have some specific profiles $\ell(r,y)$, such that, even if $\epsilon\ne0$, the isosurfaces remain symmetric with respect to $y=0$, i.e., the contour $\mathcal{W}$ satisfies $r(y)=r(-y)$. In the following, we will prove that, assuming $\ell(r,y)$ to be well-defined near $y=0$, such specific profiles of $\ell(r,y)$ do not exist as long as $\epsilon\ne0$. This means that as long as $\epsilon\ne0$, the shape of the accretion disk is inevitably asymmetric with respect to $y=0$, and it cannot be easily disguised by simply tuning the profile of specific angular momentum $\ell(r,y)$ of the disk fluid.   

Regarding the strategy of the proof, we will adopt the proof of contradiction. We will first assume that the contour of $\mathcal{W}$ is symmetry with respect to $y=0$, at least near $y=0$. Then, by solving Eq.~\eqref{dydr}, we will show that the resulting profiles of the specific angular momentum $\ell(r,y)$ inevitably become ill-defined near $y=0$.

For a given function $dy/dr$, Eq.~\eqref{dydr} is a quadratic equation for $\ell$. Therefore, one can obtain the algebraic solution for $\ell$ as
\begin{equation}
\ell=\ell(r,y)=\frac{-\mathcal{B}(r,y)\pm\sqrt{\mathcal{B}(r,y)^2-4\mathcal{A}(r,y)\mathcal{C}(r,y)}}{2\mathcal{A}(r,y)}\,,\label{lrysolve}
\end{equation}
where
\begin{align}
\mathcal{A}(r,y)&\equiv\frac{dy}{dr}\partial_yg^{\varphi\varphi}+\partial_rg^{\varphi\varphi}\,,\nonumber\\
\mathcal{B}(r,y)&\equiv-2\frac{dy}{dr}\partial_yg^{t\varphi}-2\partial_rg^{t\varphi}\,,\nonumber\\
\mathcal{C}(r,y)&\equiv\frac{dy}{dr}\partial_yg^{tt}+\partial_rg^{tt}\,,
\end{align}
are known functions when $dy/dr$ is provided. In order to have a well-defined $\ell(r,y)$ in the space region $(r,y)$ of interest, say, $y\approx0$ and outside the event horizon, the discriminant $\mathcal{D}(r,y)\equiv\mathcal{B}(r,y)^2-4\mathcal{A}(r,y)\mathcal{C}(r,y)$ should be nonnegative in that region.

To proceed, we assume that the isosurfaces of  $\mathcal{W}$ are symmetric with respect to $y=0$. More explicitly, the contour $dy/dr$ is assumed to be expressed as $dy/dr=c(r)/y$ near $y=0$, with $c(r)$ being an arbitrary function of $r$. Then, the discriminant $\mathcal{D}(r,y)$ can be expressed as
\begin{equation}
\mathcal{D}(r,y)=\frac{\mathcal{F}_0(r,y)+\epsilon\mathcal{F}_1(r,y)+\epsilon^2\mathcal{F}_2(r,y)}{\Delta(r)^2\left(1-y^2\right)^2\left(r^2+a^2y^2+\tilde\epsilon(y)\right)^4}\,,\label{discr}
\end{equation}
where $\Delta(r)\equiv r^2-2Mr+a^2$, and
\begin{widetext}
\begin{align}
\mathcal{F}_0(r,y)=&\,16 M\left(r^2 + 
     a^2 y^2\right)^2 \left[r (1 - y^2) (r^2 - a^2 y^2) + (a^2 - r^2) (r^2 + 
        a^2 y^2) c(r) - 2 a^2 r \Delta(r) c(r)^2\right]\,,\\
        \mathcal{F}_1(r,y)=&\,\frac{8aM^2}{y}\Big\{2 r y^2 \left(1- y^2\right) \left[(4M - r) r^3 - a^2 r^2 \left(1 + y^2\right) + 
     a^4 y^2 \left(1 - 2 y^2\right)\right]\nonumber\\+ \big[ 
     a^6 y^4\left(1 + 3 y^2\right)&+a^4 r^2 y^2 \left(17y^2-6 - 3 y^4\right)  - r^5 \left(3 r \left(y^2-1\right) + 4M \left(1 + y^2\right)\right) + 
     a^2 r^3 \left(4M y^2 \left(3 - 5 y^2\right) + r \left(1 + y^2 + 2 y^4\right)\right)\big] c(r)\nonumber\\&- 
  2 r \Delta(r) \left[r^4 + a^2 r^2 \left(5y^2-1\right) + 
     a^4 \left(y^2 + 2 y^4\right)\right] c(r)^2\Big\}\,,\label{F1}\\
     \mathcal{F}_2(r,y)=&\,\frac{M^3}{y^2}\Big\{16 a^4 r y^4 \left(1 - y^2\right)^2 + 
  8 y^2 \left[a^2 (4M - r) r^3 \left(1 - 3 y^2\right) - a^4 r^2 \left(3 - 8 y^2 + y^4\right) +
      a^6 \left(y^2 + y^4\right)\right] c(r)\nonumber\\&- 
  8 a^2 r \Delta(r) \left[ r^2 \left( 3 y^2-1\right) + a^2y^2 \left(1 + y^2\right)\right] c(r)^2\Big\}\,.\label{F2}
\end{align}
\end{widetext}
The denominator on the right-hand side of Eq.~\eqref{discr} is nonnegative. Therefore, whether the discriminant $\mathcal{D}(r,y)$ can become negative somewhere depends on the overall sign of the numerator.

In the limit at $r\rightarrow\infty$ and small $|y|$, the function $\mathcal{F}_2$ given by Eq.~\eqref{F2} can be approximated as
\begin{equation}
\mathcal{F}_2(r,y)\approx\frac{8a^2M^3r^5c(r)^2}{y^2}\,,
\end{equation}
while the terms containing $c(r)^2$ in $\mathcal{F}_1(r,y)$ in the same limit, from Eq.~\eqref{F1}, is given by $-16aM^2r^7c(r)^2/y$, which dominates over $\mathcal{F}_2(r,y)$ when $r$ is large. Therefore, the numerator on the right-hand side of Eq.~\eqref{discr} is dominated by $\mathcal{F}_1(r,y)$ at the limit where $r\rightarrow\infty$ and $y\rightarrow0$. We note that the term containing $\mathcal{F}_1(r,y)$ has an overall factor $\epsilon a/y$, which always goes to $-\infty$ on one side of $y=0$ despite the sign of $\epsilon a$. As a result, the discriminant $\mathcal{D}(r,y)$ always has a negative region on one side of the equatorial plane, and $\ell(r,y)$ given in Eq.~\eqref{lrysolve} is hence not well-defined within that region. Therefore, the assumption of equatorial symmetric isosurfaces of $\mathcal{W}$ leads to an ill-defined profile of $\ell(r,y)$ near the equatorial plane.

\begin{figure*}
    \centering
    \begin{tabular}{cc}
         \includegraphics[width=0.5\linewidth]{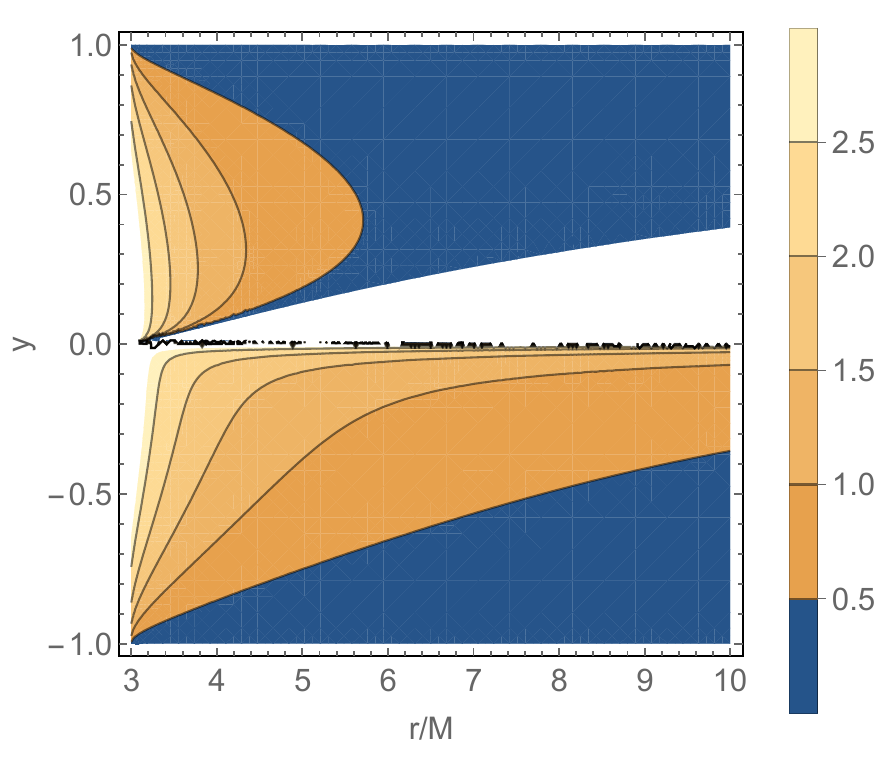}& \includegraphics[width=0.5\linewidth]{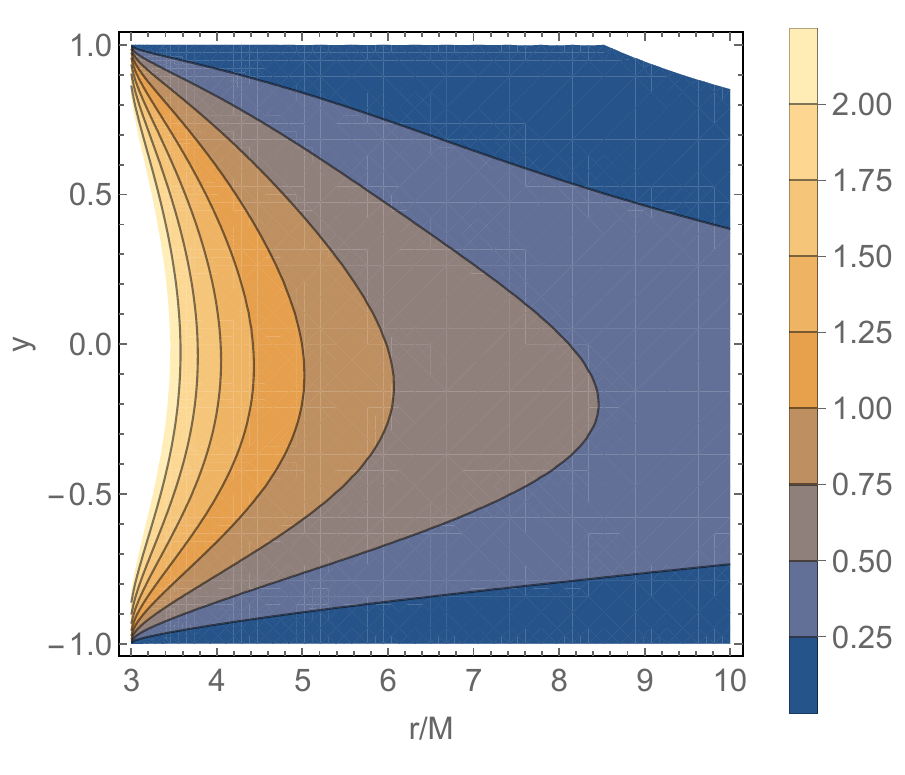} 
    \end{tabular}
    
    \caption{The contour of $\ell(r,y)$ assuming $dy/dr=(3M-r)/(M^2y)$. The left and right panels correspond to $\tilde\epsilon(y)=\epsilon aMy$ and $\tilde\epsilon(y)=\epsilon aMy^3$, respectively. We fix $a/M=1/2$ and $\epsilon=1$.}
    \label{fig:contourlry}
\end{figure*}

To demonstrate this result more explicitly, we assume that the contour $dy/dr$ of $\mathcal{W}$ is given by simple concentric circles with the center located at $r=3M$ at the equatorial plane, i.e., $dy/dr=(3M-r)/(M^2y)$. Then, by solving Eq.~\eqref{lrysolve} numerically, we show the profiles of the specific angular momentum $\ell(r,y)$ in Fig.~\ref{fig:contourlry}. In the left panel, where we consider $\tilde\epsilon(y)=\epsilon aMy$, the white region corresponds to the ill-defined region in the $\ell(r,y)$ distribution. The ill-defined region happens on one side of the equatorial plane and is precisely due to the fact that the discriminant $\mathcal{D}$ becomes negative there. In particular, the area of the white region grows as $r$ increases, which is consistent with our previous analysis in which the large-$r$ limit was taken. In the right panel, we show that a well-defined, while quite fine-tuned $\ell(r,y)$ is possible if we choose $\tilde\epsilon(y)=\epsilon aMy^3$. However, to the best of our knowledge, in other models of black holes that break the $\mathbb{Z}_2$ symmetry in the literature \cite{Cardoso:2018ptl,Cano:2019ore,Cano:2022wwo,Tahara:2023pyg}, the leading-order corrections on top of the Kerr line element are always proportional to $y$. Indeed, there is no fundamental reason to assume that these leading-order correction terms appear as $y^3$ or even higher powers. Therefore, such a choice $\tilde\epsilon(y)=\epsilon aMy^3$ may be relatively less motivated from a physical point of view.

Finally, we extend our analysis by considering a more general scenario. We would like to show that the ill-defined distribution of $\ell(r,y)$ near the standard equatorial plane, assuming a symmetric $\mathcal{W}$, is a general feature in the sense that it also happens in other black hole models that break $\mathbb{Z}_2$ and is not limited by the No$\mathbb{Z}$ model specifically considered here. To demonstrate this, we consider the following general metric components
\begin{align}
g^{tt}=g^{tt}_{\textrm{Kerr}}+yf_1(r,y^2)\,,\nonumber\\
g^{t\varphi}=g^{t\varphi}_{\textrm{Kerr}}+yf_2(r,y^2)\,,\nonumber\\
g^{\varphi\varphi}=g^{\varphi\varphi}_{\textrm{Kerr}}+yf_3(r,y^2)\,,\label{generalmetric}
\end{align}
where $f_1$, $f_2$, and $f_3$ are functions of $r$ and $y^2$, and they satisfy $f_i(r,0)\ne0$. In this model \eqref{generalmetric}, the non-Kerr correction terms that break $\mathbb{Z}_2$ symmetry, at the leading order, are assumed to be proportional to $y$. Note that in the No$\mathbb{Z}$ black hole model whose metric components are given by Eqs.~\eqref{gtt}-\eqref{g23} with $\tilde\epsilon(y)=\epsilon aMy$, the functions $f_1$, $f_2$, and $f_3$ can be explicitly expressed as
\begin{align}
f_1(r,y^2)\Big|_{\textrm{No}\mathbb{Z}}&=\frac{2a\epsilon M^2r\left(r^2+a^2\right)}{\Delta(r)\left(r^2+a^2y^2\right)^2}\,,\nonumber\\
f_2(r,y^2)\Big|_{\textrm{No}\mathbb{Z}}&=\frac{2a^2\epsilon M^2r}{\Delta(r)\left(r^2+a^2y^2\right)^2}\,,\nonumber\\
f_3(r,y^2)\Big|_{\textrm{No}\mathbb{Z}}&=-\frac{a\epsilon M}{\left(r^2+a^2y^2\right)^2}\left[\frac{1}{1-y^2}-\frac{a^2}{\Delta(r)}\right]\,.\label{f1f2f3noz}
\end{align}

Considering the general case \eqref{generalmetric} and assuming again $dy/dr=c(r)/y$, we repeat the same analysis as we did before to obtain the distribution $\ell(r,y)$ using Eq.~\eqref{lrysolve}. In this case, the discriminant $\mathcal{D}(r,y)\equiv\mathcal{B}(r,y)^2-4\mathcal{A}(r,y)\mathcal{C}(r,y)$ near $y=0$ can be approximated as
\begin{equation}
\mathcal{D}(r,y)=\frac{\mathcal{G}_{-2}(r)}{y^2}+\frac{\mathcal{G}_{-1}(r)}{y}+\mathcal{O}(y^0)\,,\label{Dgeneral}
\end{equation}
where
\begin{widetext}
\begin{align}
\mathcal{G}_{-2}(r)\equiv&\, 4c(r)^2\left[f_2(r,0)^2-f_1(r,0)f_3(r,0)\right]\,,\nonumber\\
\mathcal{G}_{-1}(r)\equiv& -\frac{8c(r)}{r^2\Delta(r)^2}\Big[f_1(r,0)\left(a^2M-4M^2r+4Mr^2-r^3\right)-f_2(r,0)\left(2a^3M-8aM^2r+6aMr^2\right)\nonumber\\
&+f_3(r,0)M\left(a^4-4a^2Mr+2a^2r^2+r^4\right)\Big]\nonumber\\
&-\frac{8c(r)^2}{r^3\Delta(r)}\left[f_1(r,0)\left(2a^2M-2M^2r+r^3\right)-4a^3Mf_2(r,0)+2a^2M f_3(r,0)\left(r^2+a^2\right)\right]\,.\label{Gm2m1}
\end{align}
\end{widetext}

We recall that, at the large-$r$ limit, the metric functions of the Kerr spacetime have the following asymptotic behaviors:
\begin{align}
g^{tt}_{\textrm{Kerr}}=&-1-\frac{2M}{r}+\mathcal{O}(r^{-2})\,,\nonumber\\g^{t\varphi}_{\textrm{Kerr}}\approx&-\frac{2aM}{r^3}\,,\quad g^{\varphi\varphi}_{\textrm{Kerr}}\approx\frac{1}{r^2\left(1-y^2\right)}\,.
\end{align}
Suppose the model reduces to the Kerr spacetime at large $r$ and the mass of the black hole is still given by $M$. We assume that the asymptotic behaviors of the functions $f_1$, $f_2$, and $f_3$ in the large-$r$ expansion are given by $f_1\approx\mathcal{O}(r^{-1-n})$, $f_2\approx\mathcal{O}(r^{-3-n})$, and $f_3\approx\mathcal{O}(r^{-2-n})$, where $n>0$. The fact that the radial fall-off behaviors of these functions are controlled by a single index $n$ can be interpreted as the assumption that the $\mathbb{Z}_2$ asymmetry of the spacetime, albeit manifesting in different components of the metric, is contributed by a single physics. We note that for the No$\mathbb{Z}$ model, i.e., Eq.~\eqref{f1f2f3noz}, the expansion of the metric components at large $r$ gives $n=2$.

Having these assumptions on the large-$r$ behaviors of $f_1$, $f_2$, and $f_3$, one can see from Eq.~\eqref{Gm2m1} that, at large $r$, $\mathcal{G}_{-2}$ is approximated as $-4c(r)^2f_1f_3$, and the terms containing $c(r)^2$ in $\mathcal{G}_{-1}$ are dominated by $-8c(r)^2f_1/r^2$. Because $f_3\approx\mathcal{O}(r^{-2-n})<\mathcal{O}\left(r^{-2}\right)$ at large $r$, the term $\mathcal{G}_{-1}/y$ in Eq.~\eqref{Dgeneral} dominates the discriminant $\mathcal{D}(r,y)$ at large $r$. Due to the $1/y$ factor in the $\mathcal{G}_{-1}/y$ term, the discriminant always becomes negative on one side of the equatorial plane $y=0$, where the distribution of $l(r,y)$ is ill-defined. We then conclude that if a black hole geometry is not symmetric with respect to the equatorial plane, the isosurfaces of the effective potential $\mathcal{W}$ are also asymmetric with respect to that plane. Any asymmetric distribution of the specific angular momentum $\ell(r,y)$ that attempts to produce a symmetric $\mathcal{W}$ either turns out to be ill-defined near the equatorial plane or suffers from fine-tuning issues, hence is not physically sound.

\section{\label{sec:con}Conclusions}

The equatorial reflection ($\mathbb{Z}_2$) symmetry is one of the fundamental spacetime symmetries of the Kerr geometry. Any observational evidence that indicates the lack of such a symmetry for isolated black holes implies the violation of the Kerr hypothesis. In order to test the Kerr hypothesis through testing the $\mathbb{Z}_2$ symmetry of black holes, it is desirable to look for possible features that can appear if and only if the $\mathbb{Z}_2$ symmetry is broken.

As has been shown in Ref.~\cite{Chen:2021ryb}, one such feature can be identified from the Keplerian thin accretion disk around the black hole. More explicitly, when $\mathbb{Z}_2$ symmetry is broken, the Keplerian thin disk generically acquires a curved surface, as opposed to the flat disk plane at the equatorial plane in the usual $\mathbb{Z}_2$ symmetric cases. In this paper, we extend the analysis of Ref.~\cite{Chen:2021ryb} by considering thick disk tori configurations described by non-self-gravitating Polish doughnut models. Employing a constant specific angular momentum profile for the moving fluid, i.e., $\ell(r,y)=\ell_0$, we identify the critical points in the tori configurations, that is, the center and the cusp. The center and the cusp correspond to the local minimum and the saddle point of the effective potential $\mathcal{W}$, respectively. Since these two points also belong to the stable and unstable branches of the Keplerian circular orbits, which are shifted away from the equatorial plane by a radius-dependent amount, the critical points, as well as the entire tori configurations, are not symmetric with respect to the equatorial plane either. In particular, in the No$\mathbb{Z}$ black hole model considered in this paper, the center, the cusp, and the entire torus configuration are twisted toward the same direction with respect to the equatorial plane. The same argument also applies to non-constant profiles of $\ell(r,y)$ if the specific angular momentum profile is symmetric with respect to the standard equatorial plane $y=0$.

Based on these results, we go further by addressing the following question: Is it possible to have a symmetric tori configuration by considering $\mathbb{Z}_2$ asymmetric specific angular momentum profiles, even if such profiles may be rather fine-tuned? By assuming that the leading-order corrections that generate $\mathbb{Z}_2$ asymmetry in the metric components are proportional to $y$ (or $\cos\theta$), we show that the specific angular momentum profiles that give rise to a symmetric tori configuration inevitably become ill-defined near the standard equatorial plane. Therefore, curved thin disks and a $\mathbb{Z}_2$ asymmetrically distorted thick tori could be a special feature that appears if and only if isolated black holes do not have $\mathbb{Z}_2$ symmetry.

Finally, we would like to emphasize a potential observational implication of the results presented in this paper. In Ref.~\cite{Chen:2024hpw}, it was shown that, by considering the No$\mathbb{Z}$ black hole spacetime surrounded by an optically thick Keplerian accretion disk, a special concave shape may appear in the approaching side of the inner part of the disk morphology for a near-edge-on observer, leaving some imprints on the line profiles. Such a concave shape emerges because the light rays and the disk configurations are shifted toward opposite directions with respect to the standard equatorial plane, such that some light trajectories at the approaching side of the disk near the black hole, which would have crossed the disk in the $\mathbb{Z}_2$ symmetric cases, would miss the disk. For the optically thick and geometrically thick tori configuration considered in this paper, the critical points, i.e., the center and cusp, as well as the entire tori structure, are shifted toward the same direction as that of the stable Keplerian circular orbits. Therefore, it is expected that the light rays and the tori configuration are shifted also toward opposite directions with respect to the equatorial plane. It is thus plausible that similar observational features as those found in Ref.~\cite{Chen:2024hpw} may also appear for accretion disks beyond geometrically thin and Keplerian. A detailed analysis along this direction will be presented elsewhere. In addition, it will be interesting to investigate whether the $\mathbb{Z}_2$ asymmetric torus configurations studied here leave measurable imprints on quasi-periodic oscillations (QPOs). Since QPO frequencies are typically associated with orbital and epicyclic frequencies of matter in the strong-gravity region, the displacement of stable circular orbits and of the torus center away from the standard equatorial plane may lead to systematic shifts in the predicted frequency spectrum. Such deviations could, in principle, modify both the values of the fundamental frequencies and the resonance conditions commonly invoked in high-frequency QPO models. A natural next step is therefore to compute the oscillation spectrum of thick tori in these backgrounds and directly fit the resulting frequency predictions to observed QPO data from X-ray binaries. This would allow one to assess whether the $\mathbb{Z}_2$ asymmetric geometry provides improved fits compared to standard reflection-symmetric spacetimes, and to derive observational constraints on the degree of symmetry breaking. Finally, to investigate the impact of a twisted accretion disk arising from a $\mathbb{Z}_2$ asymmetry on strong field observations in more realistic setups, it would be interesting to incorporate such a model in GRMHD simulations coupled to ray-tracing to simulate the consequences for EHT-like observations of the black hole shadow.

\section*{Acknowledgement}
CYC is supported by the Special Postdoctoral Researcher (SPDR) Program at RIKEN and RIKEN Incentive Research Grant (Shoreikadai) 2025. AT and EH acknowledge the Deutsche Forschungsgemeinschaft (DFG, German Research Foundation) funded under the project number 510727404, and under Germany’s Excellence Strategy-EXC-2123 QuantumFrontiers-390837967. A. T. also acknowledges the Institute for Theoretical Physics of the Leibniz University of Hannover, for support and hospitality as a guest researcher.

\bibliographystyle{utphys}
\bibliography{bib}

\end{document}